\documentclass[11pt]{article}
\usepackage{color}
\usepackage{amsmath,amssymb}
\usepackage{graphicx} 
\usepackage{bbm}
\usepackage{longtable}
\usepackage[small,loose]{subfigure}
\usepackage[small]{caption2} 
\usepackage{fleqn} 
\usepackage{cite}
\usepackage{pifont}

\advance \headheight by 3.0truept       
\setcaptionwidth{0.85\textwidth}

\DeclareMathOperator{\tr}{tr}

\newcommand{\I}{\mathrm{i}}

\newcommand{\PQ}{\ensuremath{\mathrm{U(1)}_\mathrm{PQ}}}

\newcommand{\E}[1]{\ensuremath{\mathrm{E}_{#1}}} 

\newcommand{\SO}[1]{\ensuremath{\mathrm{SO}(#1)}}
\newcommand{\SU}[1]{\ensuremath{\mathrm{SU}(#1)}}
\newcommand{\U}[1]{\ensuremath{\mathrm{U}(#1)}}
\newcommand{\Z}[1]{\ensuremath{\mathbbm{Z}_{#1}}} 

\newcommand{\bs}[1]{\ensuremath{\boldsymbol{#1}}}
\newcommand{\bsb}[1]{\ensuremath{\boldsymbol{\overline{#1}}}}
\newcommand{\maG}{\ensuremath{\mathcal{G}} }
\newcommand{\maN}{\ensuremath{\mathcal{N}} }

\allowdisplaybreaks

\begin{document}

\date{}
\title{
\begin{flushright}
\normalsize{DESY-09-021}\\
\normalsize{KUNS-2190}\\
\normalsize{LMU-ASC 08/09}\\
\end{flushright}
\vskip 2cm
{\bf\huge Accions}\\[0.8cm]}

\author{{\bf\normalsize
Kang-Sin~Choi$^1$\!,
Hans~Peter~Nilles$^2$\!,
Sa\'ul~Ramos-S\'anchez$^3$\!,}
\\{\bf\normalsize
Patrick~K.S.~Vaudrevange$^4$\!} \\[1cm]
{\it\normalsize
${}^1$ Department of Physics, Kyoto University, Kyoto 606-8502, Japan}\\[2mm]
{\it\normalsize
${}^2$ Bethe Center for Theoretical Physics and}\\[-0.5mm]
{\it\normalsize Physikalisches Institut der Universit\"at Bonn,}\\[-0.5mm]
{\it\normalsize Nussallee 12, 53115 Bonn, Germany}\\[2mm]
{\it\normalsize
${}^3$ Deutsches Elektronen-Synchrotron DESY, Hamburg, Germany}\\[2mm]
{\it\normalsize
${}^4$ Arnold Sommerfeld Center for Theoretical Physics,}\\[-0.05cm]
{\it\normalsize Ludwig-Maximilians-Universit\"at M\"unchen, 80333 M\"unchen, Germany}
}

\maketitle 

\thispagestyle{empty}

\begin{abstract}
{Axion fields provide the most elegant solution to the strong
CP problem. In string compactifications it is
difficult to obtain an axion whose decay constant is
consistent with current cosmological bounds. We examine
this question in theories with accidental \U1
symmetries that appear as low energy remnants of discrete
symmetries. We refer to the axion-like particles from the
spontaneous breakdown of these symmetries as accions.
In such systems, the accion decay constant depends
on the vacuum configuration and can be lowered to
fit the bounds. Furthermore, we find that such accions with consistent
decay constant can be embedded in special vacua of
\Z6--II orbifold models with realistic features.
}
\end{abstract}
\clearpage

\section{Introduction}

The most elegant and appealing solution to the strong CP-problem is based
on the conjecture of an axion field~\cite{Weinberg:1977ma,Wilczek:1977pj}. 
It requires the existence of an
anomalous global Peccei-Quinn symmetry~\cite{Peccei:1977hh}
\PQ\ and its
spontaneous breakdown at a scale $F_a$ (where $F_a$ denotes the ``axion
decay constant''). Constraints (mostly) from astrophysics and cosmology
require $F_a$ to be in the axion window
\begin{equation}
\label{eq:AxionBound}
10^9 \,{\rm GeV} \leq F_a \leq 10^{12} \,{\rm GeV}
\end{equation}
for the so-called ``invisible'' axion~\cite{Kim:1979if,Dine:1981rt}.
The axion field adjusts its
vacuum expectation value (VEV) to cancel the $\theta$-parameter of
quantum chromodynamics (QCD) to avoid CP-violation due to strong
interactions.

In the presence of gravity, the existence of an exact global \U1-symmetry
might be problematic. This holds explicitly in the context of
string theory, where potential anomalous \U1-symmetries appear
as gauge (not global) symmetries~\cite{Banks:1988yz}. Does this
exclude the axion solution to the strong CP-problem in the
framework of string theory?

In the present work, we focus on a set of realistic string extensions
of the minimal supersymmetric standard model (MSSM) in the framework
of the $\E8\times\E8$ heterotic string. We address the following
two questions:
\begin{itemize}
\item how can we possibly obtain a global \PQ?
\item can we disconnect the axion scale $F_a$ from the string scale
$M_s$ (which is typically much larger, e.g. $10^{16}-10^{18}$ {\rm GeV})?
\end{itemize}
The analysis of the models reveals (among the various local 
\U1-symmetries) the appearance of a multitude of discrete
symmetries of various types~\cite{Kobayashi:2006wq,Araki:2008ek}.
In such a situation we might find
accidental \U1-symmetries~\cite{Weinberg:1972fn} 
if we restrict our attention to superpotentials of a limited degree
in the polynomial of the fields. Higher order non-renormalizable
terms would break this accidental \PQ-symmetry back to the
exact discrete symmetry.~\footnote{Similar ideas
have been explored in the context of Calabi-Yau 
compactifications~\cite{Lazarides:1985bj}.}
In the standard model of particle physics (SM)
such accidental symmetries appear as $\U1_{\rm B,L}$  where B(L) denote
baryon(lepton) number, respectively.

Including these accidental symmetries will allow for a multitude of 
axion candidates in the view of the first question. Apart from the
strong CP-problem these fields might be of relevance for hidden sector
gauge groups (embedded in the second \E8) or quintessential
axions~\cite{Kim:1998kx,Kim:2002tq}.

This new paradigm of string theory, the appearance of accidental
global symmetries from exact discrete symmetries, might find interesting
applications in the creation of hierarchical structures as 
explained in~\cite{Kappl:2008ie}. This might then give us the freedom to lower
the axion scale away from the string scale and thus answering our
second question.

Within the application to the axion system, pioneering work has
been done by Choi, Kim and Kim~\cite{Choi:2006qj} in a model based on heterotic
string constructions via the \Z{12}-orbifold. Unfortunately,
however, in this model the axion decay constant of the QCD axion
could not be decoupled from the string scale. In the present paper 
we analyze the models of the so-called heterotic 
Mini-Landscape~
\cite{Lebedev:2006kn,Lebedev:2006tr,Lebedev:2007hv,Lebedev:2008un,Nilles:2008gq}
in view of potential realistic axion candidates. This requires
a careful study of a multi-axion system including a
multitude of global and discrete symmetries both of anomalous
and non-anomalous nature. Our analysis of section 2 reveals the 
fact that the properties of the system (apart from the structure
of the model itself) will strongly depend on the vacuum configuration.

In section~\ref{sec:AxionsInOrbifolds} we shall then discuss some representative examples of the
Mini-Landscape and show their potential suitability for the answers
of the questions raised above. This discussion requires extensive computer
calculations and we limit ourselves here to the conceptual questions towards
the solution of the strong CP-problem via a suitable axion field. The
construction of a complete and fully realistic extension of the MSSM will
be relegated to future investigations.

We shall see that accidental symmetries that naturally appear in the
framework of string constructions are very well suited for satisfactory
axion candidates with a decay constant in the axion window. Section~\ref{sec:conclusions}
will summarize our results, address remaining problems and will give
an outlook for future research.

\section{Multi-Axion Systems}
\label{sec:axions}

The models under consideration have, in addition to some non-anomalous $\U1_\textrm{gauge}$ 
gauge symmetries, (accidental) anomalous \U1 symmetries
whose breaking provides many axion candidates.
This leads to a complicated vacuum structure
that requires a careful investigation. Here we rely on previous work
\cite{Lebedev:2006kn,Lebedev:2006tr,Lebedev:2007hv,Lebedev:2008un,Nilles:2008gq}
where the supersymmetric vacua corresponding to D- and F-flat solutions 
have been analyzed. The models typically have one\footnote{More are
possible in the blow-up version of the models~\cite{GrootNibbelink:2007ew}} 
anomalous gauge symmetry $\U1_{\rm A}$~\cite{Witten:1984dg}
with a non-trivial Fayet-Iliopoulos (FI) term
which in turn induces non-vanishing VEVs for some of the scalar fields.
These break some of the $\U1_\textrm{gauge}$ symmetries as well as the 
$\U1_{\rm A}$ at a scale one or two orders of magnitude below the 
string scale $M_s$. We now also have to analyze the fate of the
possible accidental (anomalous) \U1 symmetries. To answer the
question towards a satisfactory axion within the ``allowed window''
we have to understand not only the model 
but also the vacuum configuration, which we 
discuss now.

\subsection{Axions  from a Single  \bs{\U1}}
\label{sec:AxionFromASingleU1s}

Let us start with the simplest case,
a supersymmetric theory with standard model (SM) gauge group 
and one additional global symmetry, \PQ, which is assumed to have a mixed 
$\PQ-\SU3_C-\SU3_C$ anomaly. Consider the matter
superfields $(\varphi_i,\psi_i)$ with \PQ\ charges $q^i$. 
If one scalar SM singlet field, say
$\varphi_0$, with non-vanishing PQ 
charge attains a VEV $v_0$, its phase
field $a$ becomes an axion and induces the following term in the 
effective Lagrangian 
\begin{equation}
  \label{eq:Leff}
  \mathcal{L}_\mathrm{eff} \supset \frac{a}{F_a} \frac{g^2}{32\pi^2}\,\tr\left[G^{\mu\nu}\widetilde{G}_{\mu\nu}\right]\,,
\end{equation}
where $g$ denotes the $\SU3_C$ coupling constant, $G_{\mu\nu}$ 
is the color field strength, and
$\widetilde G_{\mu\nu}$ its dual. The axion decay constant $F_a$ 
in this scenario is given by
\begin{equation}
  \label{eq:ADecayConstant}
  F_a ~=~ \frac{q^0\;v_0}{\mathcal{A}}
\end{equation}
that does not depend on the normalization of the PQ symmetry.
Here $\mathcal{A}$ is the coefficient of the $\PQ-\SU3_C-\SU3_C$ anomaly
\begin{equation}
  \label{eq:colorAnomaly}
  \mathcal{A} ~=~ \sum_i q^i\,\ell_i
\end{equation}
with $\ell_i$ being the $\SU3_C$ quadratic index of the fermion 
$\psi_i$, such that $\ell_i=1$ for $\psi_i$
transforming as a triplet. Note that the chiral PQ transformation 
of the axion,
$a\rightarrow a + q^0\,v_0\,\alpha$ (with $\alpha$ being the 
transformation parameter), reproduces as
expected the original anomaly,
\begin{equation}
  \label{eq:LeffAnomaly}
  \mathcal{L}_\mathrm{eff} \rightarrow \mathcal{L}_\mathrm{eff} 
         + \alpha\mathcal{A}\,\frac{g^2}{32\pi^2}\,\tr\left[G^{\mu\nu}\widetilde{G}_{\mu\nu}\right]\,.
\end{equation}

It is straightforward to generalize this result to the case where various
scalar fields obtain a VEV.
If the fields $\varphi_i$ attain VEVs
$v_i$, the resulting axion $a$ is given in terms of the phase fields $a_i$ by
\begin{equation}
  \label{eq:AxionNFields}
  a ~=~ \frac{1}{M} \left(\sum_i a_i\,q^i\,v_i\right)
    ~\equiv~\frac{1}{\sqrt{\sum_i (q^i\,v_i)^2}} \left(\sum_i a_i\,q^i\,v_i\right)\,,
\end{equation}
where $M$ defines the scale at which \PQ\ is broken. The axion decay constant is then given by
\begin{equation}
  \label{eq:ADecayConstantNFields}
    F_a ~=~ \frac{M}{\mathcal{A}}\,.
\end{equation}
Considering the PQ transformation of the phase fields 
$a_i$, one recovers eq.~\eqref{eq:LeffAnomaly}. We
would like to point out that in this case the axion decay 
constant is dominated by the largest VEV. 
This
behavior might change if the low energy theory contains more 
than one PQ symmetry, as we shall discuss below.

\subsection{Axions  from Multiple \bs{\U1}s}
\label{sec:AxionFromMultipleU1s}

In a next step we consider a supersymmetric theory with two 
PQ symmetries, denoted by $\U1_P\times\U1_Q$, with respective
(non-vanishing) anomalies $\mathcal{A}_P$ and $\mathcal{A}_Q$, 
analogously defined as in eq.~\eqref{eq:colorAnomaly}. 
The scalar fields of the theory transform as
\begin{equation}
  \label{eq:TrafoScalar2U1}
   \varphi_i \rightarrow e^{\I\left(\alpha\,q_P^i+ \beta\,q_Q^i\right)} \varphi_i
\end{equation}
under these symmetries. Note that the \U1 resulting from the linear combination
\begin{equation}
  \label{eq:AfreeCharge}
  q_f^i ~=~ \mathcal{A}_Q\,q_P^i - \mathcal{A}_P\, q_Q^i
\end{equation}
is always anomaly-free.

Let us suppose that both \U1s are broken spontaneously. 
This occurs if, for instance, two fields $\varphi_1$ and
$\varphi_2$ develop VEVs, provided that their charges 
$(q_P^1,\,q_Q^1)$ and $(q_P^2,\,q_Q^2)$ 
are linearly independent,
i.e.
\begin{equation}
  \label{eq:ChargesCondition}
  q_P^1\,q_Q^2~\neq~ q_Q^1\,q_P^2\,.
\end{equation}
If the charges do not satisfy eq.~\eqref{eq:ChargesCondition}, 
only one \U1 would
be broken. As before, the spontaneous breaking of these \U1s induces the term 
\begin{equation}
  \label{eq:Leff2U1}
  \mathcal{L}_\mathrm{eff}~=~\frac{a}{F_a} \frac{g^2}{32\pi^2}\,\tr\left[G^{\mu\nu}\widetilde{G}_{\mu\nu}\right]
     ~\equiv~\left(\frac{a_1}{F_{a_1}} + \frac{a_2}{F_{a_2}}\right)
              \frac{g^2}{32\pi^2}\,\tr\left[G^{\mu\nu}\widetilde{G}_{\mu\nu}\right]\,.
\end{equation}
The constants $F_{a_i}$ are found by comparing the variation of $\mathcal{L}_\mathrm{eff}$ due to the
chiral PQ transformation of the phase fields, $a_i\,\rightarrow\,a_i+v_i\,(q^i_P\,\alpha+q^i_Q\,\beta)$, and
the one expected from the anomalous nature of $\U1_P\times\U1_Q$,
\begin{equation}
  \label{eq:LeffAnomaly2U1}
  \mathcal{L}_\mathrm{eff} \rightarrow \mathcal{L}_\mathrm{eff} 
         + \left(\alpha\mathcal{A}_P +\beta \mathcal{A}_Q\right)
         \,\frac{g^2}{32\pi^2}\,\tr\left[G^{\mu\nu}\widetilde{G}_{\mu\nu}\right]\,.
\end{equation}
The solutions are
\begin{equation}
  \label{eq:AxionDecayConstants2U1}
  F_{a_1}~=~ -v_1\; \frac{q_P^1\,q_Q^2-q_Q^1\,q_P^2}{q_f^2}\,,
 \qquad\qquad
  F_{a_2}~=~ v_2\; \frac{q_P^1\,q_Q^2-q_Q^1\,q_P^2}{q_f^1}\,.
\end{equation}
Note that these constants do not depend on the basis chosen 
for the \U1s. Then, the axion and its decay
constant are respectively given by 
\begin{subequations}
\label{eq:AxionResults2U1}
\begin{eqnarray}
  \label{eq:Axion2U1}
  a & = & \frac{-a_1\,q^2_f\,v_2 + a_2\,q^1_f\,v_1}{\sqrt{(q_f^1\,v_1)^2 + (q_f^2\,v_2)^2}}\,,\\
  \label{eq:AxionDecayConstant2U1}
  F_a &=& \left( \left(\frac{1}{F_{a_1}}\right)^2 + \left(\frac{1}{F_{a_2}}\right)^2 \right)^{-1/2}
                ~=~ \frac{v_1\,v_2\,(q_P^1\,q_Q^2-q_Q^1\,q_P^2)}{\sqrt{(q_f^1\,v_1)^2 + (q_f^2\,v_2)^2}}\,.
\end{eqnarray}
\end{subequations}

A few comments are in order. First, one can easily verify that, 
contrary to the previous case, if two
\U1s are broken by two fields attaining VEVs, the axion decay 
constant given by
eq.~\eqref{eq:AxionDecayConstant2U1} is dominated by the smaller
VEV: this is the key observation in view of a realization of a
decay constant within the allowed ``axion window''. 
Secondly, apart from the axion
$a$, clearly the theory must have a Goldstone boson. It is 
the combination orthogonal to the axion, 
\begin{equation}
  \label{eq:Goldstone2U1}
  h ~=~ \frac{a_1\,q^1_f\,v_1 + a_2\,q^2_f\,v_2}{\sqrt{(q_f^1\,v_1)^2 + (q_f^2\,v_2)^2}}\,,
\end{equation}
associated to the broken anomaly-free symmetry, $\U1_f$. 
If this symmetry is gauged, $h$ is `eaten' via
the Higgs mechanism by the gauge boson of $\U1_f$, making 
it massive. If $\U1_f$ is global, $h$ remains
as a massless (uncharged) field.  As a last remark, from 
the above discussion we see that in
theories with multiple spontaneously broken \U1s, one could 
readily identify the axion as the unique
combination of the phase fields that is orthogonal to all 
Goldstone bosons. This observation will help us
in the following discussion.

\subsection{The general case}
\label{sec:GeneralCase}

The above two examples are sufficient to obtain an understanding
of the general situation, where we have more than two 
\U1 symmetries
and where the number of SM singlets acquiring 
VEVs exceeds the number of symmetries that
are broken. In order to shed some light on the form of 
the axion and its decay constant, let us examine
the scenario in which $\U1_P\times\U1_Q$ gets broken when 
three fields attain VEVs. One combination of
the three phase fields $a_i$ will provide the CP axion $a$, 
whereas the other two orthogonal combinations
will yield a Goldstone boson $h$ that does perceive chiral 
transformations, and an invariant phase field
$a'$. 

To find the axion, we can proceed as before and 
determine the constants $F_{a_i}$. A more manageable
approach is to identify first the fields $a'$ and $h$. 
The axion will be then the combination of $a_i$
orthogonal to them. 
The result is a generalization of eqs.~\eqref{eq:AxionResults2U1}. In particular,
assuming a large hierarchy of VEVs, for 
instance $v_1 \gg v_2 \gg v_3$, the axion decay constant becomes
\begin{equation}
  \label{eq:AxionDecayConstant2U13F}
  F_a\;\approx\;v_2\; \frac{q_P^1q_Q^2-q_Q^1q_P^2}{q_f^1}\,,
\end{equation}
that is, $F_a$ is dominated
by the second largest VEV ($v_2$ in our example) and neither 
by the largest nor by the smallest. The
second largest VEV corresponds to the largest scale at which 
both symmetries, $\U1_P$ and $\U1_Q$, are
already broken. So, it seems natural to expect that, for any 
number of SM fields attaining VEVs, the
axion decay constant will be dominated by the scale at which 
all PQ symmetries get broken. In the current
case, with two PQ symmetries and a large VEV
hierarchy, the scale of $F_a$ will be
given by the second largest VEV.

We can now state the result for the general case of $M$ \U1
symmetries and $N\geq M$ SM singlet fields that obtain a VEV.
By Goldstone's theorem, we have generically $M$ ``Goldstone bosons'' from the
$M$ spontaneously broken symmetries. 
This means that we have $N-M$ invariant combinations of phase fields, which
are orthogonal to these Goldstone bosons. Suppose that
one of the Goldstone bosons lies in the anomalous direction and becomes
an axion, while the other $M-1$ ones remain massless, possibly eaten
by gauge bosons. From the previous discussion, 
if the \U1s are broken at hierarchically different scales, then
the axion decay constant $F_a$ is {\em dominated by the $M$th largest
VEV}.

\subsection{Embedding in the heterotic string theory}
\label{sec:EmbeddingInStrings}

In the heterotic string, there are various sources for axions. The spontaneous
breakdown of the pseudo-anomalous $\U1_{\rm A}$ gauge symmetry together with
the Green-Schwarz mechanism produce the so-called model-independent axion,
whose decay constant is fixed by the Planck scale (see e.g.~\cite{Svrcek:2006yi}),
too high to solve the strong CP problem. Model-dependent axions arise from
the internal components of the $B-$field. Unfortunately, their decay constants
are in general not much lower than in the previous case~\cite{Svrcek:2006yi}.
Admissible axions, on the other hand, might
arise from the breaking of accidental global \U1 symmetries realized 
as low energy remnants of (stringy) discrete symmetries. In this case, 
the axion decay constant shall depend on the VEVs of the fields responsible
for the breakdown of such global \U1s.

In general, there is a set of fields with a large VEV dictated by the
FI term (one or two orders of magnitude below the
string scale). This is the scale where (some of) the gauge
\U1 symmetries are broken and where also some of the anomalous
\U1s will be broken. The latter ones should not have a
QCD anomaly (they could have e.g. hidden sector anomalies) and/or should be 
candidates for a quintessential axion. There could in addition
be another set of singlet fields that obtain smaller VEVs that
break the relevant PQ-symmetry at a scale in the axion window
and thus give a hierarchically smaller $F_a$. The requirements
for a satisfactory model are thus clear:
\begin{itemize}
\item find a model with an accidental (color)-anomalous $\U1^*$,

\item identify a vacuum configuration where the VEVs driven
by the FI term do not break $\U1^*$,

\item search for a vacuum configuration where $\U1^*$ is broken
by a VEV in the axion window (some other gauge \U1s might be broken
here as well),

\item check that higher order non-renormalizable terms that break
$\U1^*$ explicitly are sufficiently suppressed to avoid a too ``large''
axion mass.

\end{itemize}

A comment regarding the latter requirement is in order. 
Higher order terms in the superpotential break explicitly the accidental 
$\U1^*$ symmetry. This can pull the axion off from the CP-conserving vacuum and
generate an additional contribution to the axion mass 
$m_*^2$~\cite{Barr:1992qq,Kamionkowski:1992mf}.
$m_*^2$ is constrained by the bounds on the electric dipole 
moment of the neutron~\cite{Baker:2006ts,Kim:2008hd}:
\begin{equation}
  \label{eq:constraintOnmbreaking}
  m_*^2 \lesssim 10^{-11}\ m_{QCD}^2 \approx 10^{-15}\ \frac{{\rm GeV}^4}{F_a^2}\,,
\end{equation}
where $m_{QCD}^2\approx 10^{-4}\ {\rm GeV}^4 / F_a^2$ is the axion mass 
induced by QCD instantons.  
Not every $\U1^*$--violating superpotential term, however, 
contributes to the axion mass.
The magnitude of the axion mass, induced by the explicit breaking 
of the $\U1^*$ symmetry, does not necessarily depend on the order of
the lowest term in the  superpotential that breaks this symmetry. 
It rather depends 
on the order of the (lowest) $\U1^*$--violating term 
that contributes to the axion mass directly. 
Note e.g. that if a superpotential term breaks $\U1^*$ explicitly,
but contains two or more fields that do not develop VEVs, it
does not contribute to the axion mass. It follows then that verifying whether 
eq.~\eqref{eq:constraintOnmbreaking} is satisfied depends strongly on the 
specifics of the vacuum configuration and must be analyzed case by case.

We shall explore some candidate models in the next section.

\section{QCD Axions in Orbifold Models}
\label{sec:AxionsInOrbifolds}

Let us turn now to the question of whether we can find suitable QCD axions in heterotic orbifold
models. Recent progress in this framework has revealed that models resembling many features of the MSSM
can be constructed~\cite{Forste:2004ie,Kobayashi:2004ya,Buchmuller:2004hv,Buchmuller:2006ik,Kim:2006hw}. 
In particular, we analyze a subset of the models found in~\cite{Lebedev:2006kn} and identify those with the following
properties: a) their superpotentials $\mathcal{W}$ must be symmetric under an accidental PQ symmetry, 
and b) they should admit vacua such that the PQ symmetry gets spontaneously broken at a 
phenomenologically admissible scale 
(see eq.~\eqref{eq:AxionBound}).

Identifying PQ symmetries of the superpotential $\mathcal{W}$ of a model requires to verify whether all
superpotential terms of a given order preserve some \U1 symmetry other than the gauged ones.  To find
these accidental symmetries, it is convenient to use the method of~\cite{Choi:2006qj}, which we briefly
summarize in the following. The superpotential of a model can be written as
\begin{equation}
  \label{eq:W}
 \mathcal{W}=\sum_i^{\text{\# terms}} \mathcal{W}_i
    \qquad\text{ with }\qquad
 \mathcal{W}_i = \prod_j^{\text{\# fields}} \Phi_j^{\lambda_{ij}}\,,
\end{equation}
where all coefficients are set to one and $\lambda_{ij}$ are non-negative integers constrained by the order $n$ of the
superpotential according to $\sum_j \lambda_{ij}\leq n$, for all $i$. Each element of the kernel (or null
space) of the matrix $\left(\lambda^T\lambda\right)$ defines the charges of the fields under an Abelian
symmetry of $\mathcal{W}$. Subtracting those symmetries contained in the gauge group, one is left with all
accidental symmetries denoted by $\U1_\mathrm{acc}$. 

Not every accidental symmetry of the theory can play the role of a \PQ. It is crucial that such
symmetries also produce mixed $\U1_\mathrm{acc}-\SU3-\SU3$ anomalies, that is,
\begin{equation}
  \label{eq:AccidentalAnomaly}
  \mathcal{A} = \sum_j \ell_j \;q_\mathrm{acc}^j \neq 0\,,
\end{equation}
where $q_\mathrm{acc}^j$ denote the accidental charges of the fields $\varphi_j$ and $\ell_j$
refer, as before, to the quadratic indices of the \SU3 representations.

Using the method outlined above, we search for the accidental symmetries in a subset of realistic \Z6--II
orbifold models found previously in the so-called
Mini-Landscape~\cite{Lebedev:2006kn,Lebedev:2007hv,Lebedev:2008un}. We use the notation introduced  
there. We analyze 55 orbifold models based on the Mini-Landscape shift $V^{\SO{10},1}$ 
exhibiting the following properties:

 $\bullet$ SM gauge group,

 $\bullet$ non-anomalous hypercharge $\U1_Y\subset\SU5$,

 $\bullet$ spectrum $=$ 3 generations $+$ vector-like exotics,

 $\bullet$ heavy top, and

 $\bullet$ supersymmetric vacua.

The last property deserves a comment. The pseudo-anomalous symmetry $\U1_{\rm A}$ (generically
present in orbifold models) induces the FI term~\cite{Atick:1987gy}
\begin{equation}
  \label{eq:FIDTerm}
  D_\mathrm{A} ~\approx~ \sum_j q_\mathrm{A}^j\left|\langle\varphi_j\rangle\right|^2
                   + \frac{M^2_\mathrm{Pl}}{192\pi^2}\sum_j q_\mathrm{A}^j\,.
\end{equation}
Therefore, in supersymmetric vacua a set of fields $\varphi_j$ have to acquire non-vanishing VEVs, such
that $D_\mathrm{A}=0$. In general, solutions of the $D=0$ conditions can be expressed by
gauge invariant monomials~\cite{Buccella:1982nx},
\begin{equation}
  \label{eq:Monomial}
  I ~=~ \varphi_1^{m_1} \varphi_2^{m_2}\ldots\,,
\end{equation}
where $m_i$ are positive integers.
In the case of an anomalous \U1, such a monomial must be invariant only with respect to the non-anomalous
gauge symmetries in order to cancel the FI term, eq.~\eqref{eq:FIDTerm}. Since in our conventions
$\sum_j q_\mathrm{A}^j > 0$, a suitable vacuum 
configuration exists if there is a set of SM singlets, ${s_i}\subset{\varphi_j}$, that build a monomial
with a total negative anomalous charge, $\sum_j m_j\; q_\mathrm{A}^j <
0$~\cite{Buchmuller:2006ik}. Therefore, the condition $D_\mathrm{A}=0$ sets the VEVs of the singlets 
$s_i$ and, thereby, the scale of the breakdown of $\U1_{\rm A}$. In heterotic orbifolds, this scale
can be as high as $\sim10^{17}$ GeV. All models studied here have vacua satisfying $D=0$
and preserving the SM gauge group unbroken (including hypercharge).~\footnote{As we only know the form of
the superpotential without coefficients, we have to assume that $F=0$ can be fulfilled and stabilizes
the remaining $D$-flat directions. For generic coefficients of order 1, this has been checked in some
examples.}  This contradicts recent statements about model building in heterotic string
orbifolds~\cite{Tatar:2008zj}. 

For each model, we choose a monomial suitable to cancel the FI term that involves only a few SM
singlets. The VEVs of the fields in such a monomial is set close to the string scale by
eq.~\eqref{eq:FIDTerm}. Further VEVs for other SM singlets which we will consider later are assumed to
be significantly smaller. The global \U1s in which we are interested are those of the effective
superpotential at finite orders obtained after the cancelation of the FI term.
We find that the studied orbifold models have generically one or more accidental symmetries. As
presented in  table~\ref{tab:PQStatistics}, although most of the models have approximate global
symmetries up to high orders, 
only at orders 3 and 4 most of them exhibit $\U1_\mathrm{acc}-\SU3-\SU3$ anomalies. In fact,
at order 6  in $\mathcal{W}$, no model of the limited sample analyzed in this study has symmetries
which can serve as a \PQ.

\begin{table}[t!]
   \centering
   \begin{tabular}{|c|cc|}
     \hline
     Order of $\mathcal{W}$ & \# Models with some & \# Models with some\\
                            & $\U1_\mathrm{acc}$ &   \PQ\\
     \hline
     \hline
     3 & 55 & 55 \\
     4 & 55 & 54 \\
     5 & 55 & 15 \\
     6 & 34 & 0 \\
     \hline
   \end{tabular}
   \caption{Statistics of frequency of accidental PQ symmetries in a subset of promising \Z6--II heterotic orbifold
     models.} 
   \label{tab:PQStatistics}
\end{table}

The explicit breaking of all accidental PQ symmetries at order 6
in these models could spoil the CP-conserving vacuum.
However, from the discussion of sec.~\ref{sec:EmbeddingInStrings}, 
a solution to the strong CP problem is still possible provided that
the PQ-violating couplings contributing to the axion mass appear 
at higher orders. In addition, we have 
seen that requiring a hierarchy of VEVs lets us set the axion decay constant 
at an intermediate scale. Such a hierarchy could arise in this type of models 
from other symmetries of the low energy theory~\cite{Kappl:2008ie}.
In the following we examine a model with only one \PQ.

\subsection{A \Z6--II Model with a QCD Axion}
\label{sec:Example}

A \Z6--II orbifold model which can lead to a suitable QCD axion is defined by the following gauge shift
and Wilson lines: 
{\small
\begin{eqnarray}
V^{\SO{10},1}&=&\left(\frac{1}{3}, -\frac{1}{2}, -\frac{1}{2},  0,  0,  0,  0,  0\right)\left( \frac{1}{2},  -\frac{1}{6},  -\frac{1}{2},  -\frac{1}{2},  -\frac{1}{2},  -\frac{1}{2},  -\frac{1}{2},  \frac{1}{2}\right)\,,\nonumber\\
W_3&=&\left(-\frac{1}{2},  -\frac{1}{2},  \frac{1}{6},  \frac{1}{6},  \frac{1}{6},  \frac{1}{6},  \frac{1}{6},  \frac{1}{6}\right)\left(  1,  0,  -\frac{1}{3},  -\frac{2}{3},  \frac{2}{3},  -\frac{5}{3},  -\frac{2}{3},  \frac{1}{3}\right)\,,\nonumber\\
W_2&=&\left(\frac{3}{4},  \frac{1}{4},  \frac{1}{4},  \frac{1}{4},  \frac{1}{4},  -\frac{1}{4},  -\frac{1}{4},  -\frac{1}{4}\right) \left( \frac{5}{2},  -\frac{3}{2},  -2,  -\frac{5}{2},  -\frac{5}{2},  -2,  -2,  2\right)\,.
\end{eqnarray}
}
The unbroken gauge group after compactification is
\begin{equation}
  \label{eq:GG}
  \maG ~=~ \SU3_C\times\SU2_L\times\U1_Y\times[\SU4\times\SU2\times\U1_{\rm A}\times\U1^7]\,,
\end{equation}
where $\U1_Y$ denotes the standard \SU5 hypercharge and $\U1_\textrm{A}$, the anomalous \U1.
The massless matter spectrum is given in table~\ref{tab:spectrum}; it contains three MSSM
generations plus vector-like exotics with respect to $\SU3_C\times\SU2_L\times\U1_Y$. All non-Abelian and
Abelian charges of the massless particles are provided at our web page~\cite{Choi:2009xx}. 

\begin{table}[!t!]
\begin{minipage}{0.7\textwidth}
\begin{tabular}{|rlc||rlc|}
\hline
  \#  &  {\small Representation}  & {\footnotesize Label}     &  \#  & {\small (Anti-)Repr.} & {\footnotesize Label} \\
\hline\hline
3 & $( {\bf 3},  {\bf 2};  {\bf 1},  {\bf 1})_{1/6}$  & $q_i$       &    &   & \\
9 & $(  {\bf 1},  {\bf 2};  {\bf 1},  {\bf 1})_{-1/2}$ & $\ell_i$   &  6 & $(  {\bf 1},  {\bf 2};  {\bf 1},  {\bf 1})_{1/2}$  &  $\bar{\ell}_i$\\
3 & $( {\bf 1},  {\bf 1};  {\bf 1},  {\bf 1})_{1}$    & $\bar{e}_i$ &    &   & \\
3 & $( \bsb{3},   {\bf 1};  {\bf 1},  {\bf 1})_{-2/3}$ & $\bar{u}_i$ &    &   & \\
8 & $( \bsb{3},   {\bf 1};  {\bf 1},  {\bf 1})_{1/3}$  & $\bar{d}_i$ &  5 & $(  {\bf 3},  {\bf 1};  {\bf 1},  {\bf 1})_{-1/3}$ & $d_i$  \\
\hline
12& $( {\bf 1},  {\bf 1};  {\bf 1},  {\bf 1})_{1/2}$  & $s^+_i$     & 12 & $(  {\bf 1},  {\bf 1};  {\bf 1},  {\bf 1})_{-1/2}$ & $s^-_i$   \\
2 & $( {\bf 1},  {\bf 1};  {\bf 1},  {\bf 2})_{1/2}$  & $x^+_i$     &  2 & $(  {\bf 1},  {\bf 1};  {\bf 1},  {\bf 2})_{-1/2}$  & $x^-_i$ \\
\hline
\end{tabular}
\end{minipage}
\hskip -5mm
\begin{minipage}{0.25\textwidth}
\begin{tabular}{|rlc|}
\hline
  \#  &  {\small Representation}           & {\footnotesize Label}   \\
\hline\hline
 4 & $(  {\bf 1},  {\bf 2};  {\bf 1},  {\bf 1})_{0}$ & $m_i$\\
 2 & $(  {\bf 1},  {\bf 2};  {\bf 1},  {\bf 2})_{0}$ & $n_i$\\
65 & $(  {\bf 1},  {\bf 1};  {\bf 1},  {\bf 1})_{0}$ & $s_i^0$\\
12 & $(  {\bf 1},  {\bf 1};  {\bf 1},  {\bf 2})_{0}$ & $h_i$\\
10 & $(  {\bf 1},  {\bf 1};  {\bf 4},  {\bf 1})_{0}$ & $w_i$\\
10 & $(  {\bf 1},  {\bf 1};  {\bsb{4}},{\bf 1})_{0}$ & $\bar{w}_i$\\
\hline
\end{tabular}
\vskip 0.5cm\phantom{.}
\end{minipage}
\caption{Massless spectrum. The quantum numbers are shown with respect to $\SU3_C\times \SU2_L \times
  \SU4 \times \SU2$, the hypercharge is given by the subscript.}
\label{tab:spectrum}
\end{table}

The selected vacuum has to preserve $\maN=1$ SUSY. With this purpose, we require that the fields contained in
the gauge invariant monomial
\begin{equation}
  \label{eq:FIMonom}
  I_\mathrm{FI} = \left(s^0_{19}\right)^2 s^0_{49} \left(s^0_{52}\right)^3 s^0_{57}\left(s^0_{59}\right)^2
\end{equation}
(with total $\U1_{\rm A}$ charge $q_\textrm{A}(I_\mathrm{FI}) = -25/6$) acquire VEVs such that
the FI term is canceled. As a consequence, the gauge symmetry is further broken to
\begin{equation}
  \label{eq:GGAfterVEVs}
  \maG' ~=~ \SU3_C\times\SU2_L\times\U1_Y\times[\SU4\times\SU2\times\U1^3]\,.
\end{equation}

After symmetry breaking, the effective superpotential $\mathcal{W}$ of this model contains 870 couplings
of order up to five and is invariant with respect to  19 accidental \U1s, in addition to the gauge
symmetry group $\maG'$. Only one accidental \U1, that we denote \PQ, presents mixed PQ-$\SU3_C$
anomalies:  
\begin{equation}
  \label{eq:AnomalyExample}
  \mathcal{A} = \sum_i \ell_i q_\mathrm{PQ}^i = 1 \neq 0\,.
\end{equation}
If the fields entering the gauge invariant monomial
\begin{equation}
  \label{eq:PQMonom}
  I ~=~ s^0_{1}\; s^0_{36}\; s^0_{37}
\end{equation}
acquire VEVs, \PQ\ is broken spontaneously and the axion
\begin{equation}
  \label{eq:ExampleAxion}
  a = \frac{1}{M} \left(q^1_\mathrm{PQ} v_1 a_1 
              + q^{36}_\mathrm{PQ}v_{36} a_{36} 
              + q^{37}_\mathrm{PQ}v_{37} a_{37}\right)
\end{equation}
appears in the theory. Here $a_i$, $q^i_\mathrm{PQ}$ and $v_i$ denote respectively the
phase components, PQ charges and VEVs of the singlets of eq.~\eqref{eq:PQMonom}. The \PQ\ breakdown scale
$M$ is read off from eq.~\eqref{eq:AxionNFields}. 

Along the D-flat direction described by eq.~\eqref{eq:PQMonom}, the VEVs of the singlets are equal.
Thus, the axion decay constant is given by
\begin{equation}
  \label{eq:ExampleAxionCoupling}
  F_a ~=~\frac{M}{\mathcal{A}} ~=~ \sqrt{\left(q_\mathrm{PQ}^{1} v_1\right)^2 
          + \left(q_\mathrm{PQ}^{36} v_{36}\right)^2
          + \left(q_\mathrm{PQ}^{37} v_{37}\right)^2} =
          \sqrt{5}\;v_{1}\,,
\end{equation}
where we have considered the singlet charges displayed in table~\ref{tab:Singlets}.
Provided that the VEVs $v_i$ are well below the scale of the FI term, say at $\sim10^{12}$ GeV,
the inverse axion coupling can agree with the current experimental bounds, eq.~\eqref{eq:AxionBound}.
Note that we have disregarded the axion-like components of the fields that acquire VEVs
at the larger scale (see eq.~\eqref{eq:FIMonom}) as their contributions
are subdominant.

\begin{table}[t!]
  \centering
  \begin{tabular}{|l|rrrrrrrrr|r|}\hline
  Label & $q_\mathrm{A}$ & $q_{Y}$ & $q_{3}$ & $q_{4}$ & $q_{5}$ & $q_{6}$ & $q_{7}$ & $q_{8}$ & $q_{9}$ & $q_\mathrm{PQ}$ \\
  \hline
  \hline
 $s^0_{1}$  &  $\frac{5}{3 }$ & $0$ & $-1$ & $-3$ & $0$ & $0$ & $-10$ & $0$ & $0$ &  $-2$ \\
 $s^0_{19}$ & $-\frac{11}{18 }$ & $0$ & $\frac{1}{2 }$ & $\frac{11}{6 }$ & $-\frac{5}{3 }$ & $-\frac{1}{2 }$ & $-\frac{1}{2 }$ & $\frac{1}{3 }$ & $-\frac{2}{3 }$ &  $0$ \\
 $s^0_{36}$ & $\frac{22}{9 }$ & $0$ & $0$ & $4$ & $0$ & $0$ & $2$ & $0$ & $0$ &  $0$\\
 $s^0_{37}$ & $\frac{13}{9 }$ & $0$ & $1$ & $-1$ & $0$ & $0$ & $8$ & $0$ & $0$ &  $1$ \\
 $s^0_{49}$ & $\frac{7}{9 }$ & $0$ & $-\frac{1}{2 }$ & $\frac{4}{3 }$ & $\frac{5}{6 }$ & $0$ & $12$ & $-\frac{2}{3 }$ & $\frac{4}{3 }$ &  $0$\\
 $s^0_{52}$ & $-\frac{22}{9 }$ & $0$ & $0$ & $-4$ & $0$ & $0$ & $-2$ & $0$ & $0$ &  $0$ \\
 $s^0_{57}$ & $-\frac{4}{9 }$ & $0$ & $\frac{1}{2 }$ & $\frac{1}{3 }$ & $\frac{5}{6 }$ & $0$ & $-14$ & $-\frac{2}{3 }$ & $\frac{4}{3 }$ &  $0$\\
 $s^0_{59}$ & $-\frac{13}{9 }$ & $0$ & $-\frac{1}{2 }$ & $\frac{10}{3 }$ & $\frac{5}{6 }$ & $\frac{1}{2 }$ & $\frac{9}{2 }$ & $\frac{1}{3 }$ & $-\frac{2}{3 }$ &  $0$\\
 \hline
  \end{tabular}
  \caption{\U1 charges of the relevant non-Abelian singlets of a \Z6--II orbifold
   model with a PQ symmetry.}
  \label{tab:Singlets}
\end{table}

There are two issues that still have to be addressed. 
First, since the \PQ\ is broken at order 6, 
we have to verify whether there appear large contributions to the axion mass.
In the current model, the first contribution to the axion mass from the explicit breaking of \PQ\
arises at order 16. We have estimated the contribution to the axion mass to be 
$m_*^2\sim 10^{-18}\ {\rm GeV}^4 / F_a^2$, consistent with the current experimental bound, 
eq.~\eqref{eq:constraintOnmbreaking}.
Second, in the vacuum configuration chosen here, most of the exotics remain massless. This problem is due to the
small number of fields that are chosen to develop large VEVs (see eq.~\eqref{eq:FIMonom}). 
However, as shown in~\cite{Lebedev:2006kn,Lebedev:2006tr,Lebedev:2007hv,Lebedev:2008un,Nilles:2008gq}, 
there exist supersymmetric vacua where all exotics acquire masses because
(almost) all the SM singlets attain VEVs. In the current model, we find a monomial
$I_{\rm FI}$ containing 32 fields that leaves a \PQ\ symmetry unbroken and provides masses for many 
more exotics. Whether there exist configurations where all exotics are decoupled and simultaneously a
\PQ\ is retained is a question beyond the scope of the current letter and will be addressed elsewhere.
Note that this is related to the magnitude of $m_*^2$, as integrating out heavy exotics could
induce further contributions to the axion mass.

\section{Conclusions}
\label{sec:conclusions}

Discrete symmetries (as found abundantly in models of 
the heterotic brane world)
are suited to solve the strong CP problem in a satisfactory way. They
might give rise to accidental global \PQ\ symmetries as
``low energy'' remnants. Accions -- the axion like particles of the
spontaneous breakdown of these symmetries -- could have a decay 
constant $F_a$ in the allowed axion window, eq.~\eqref{eq:AxionBound}.

The key observation for this latter fact has been explained in
section~\ref{sec:axions}, formula~\eqref{eq:AxionResults2U1}. 
In a system with an anomalous and a
non-anomalous \U1 symmetry broken by two fields with nontrivial
VEVs, it is the smallest of these VEVs that corresponds to the
accion decay constant $F_a$. This allows a decoupling of $F_a$ from
the large VEVs required to cancel the FI terms (which appear to be
only one or two orders of magnitude below the string scale). The
actual value of $F_a$ depends strongly on the vacuum configuration
of the system and requires further model building. In this framework
of models with accidental \U1 symmetries, such a hierarchy of the
scales could be potentially created along the lines explained in 
ref.~\cite{Kappl:2008ie}. 

The implementation of the accion-system in the framework of realistic
MSSM models from the heterotic string is still in its infancy. The 
identification of suitable candidates for accidental \U1's
requires an enormous amount of computer work. 
In section~\ref{sec:AxionsInOrbifolds} we have examined a few 
candidate models from our previous 
MSSM search that can serve as an existence proof for the accion
solution to the strong CP-problem. A few things remain to be done: 
the decoupling of exotics at higher order in the superpotential 
and a detailed discussion of the accion mass arising from the
explicit breakdown of the accidental \U1 symmetry. Such an
analysis is beyond the scope of the present paper. It would require
substantially higher computer power and/or new algorithms to
efficiently attack the problem. Work along this direction is 
under way.

\section*{Acknowledgments}

We would like to thank Jihn E. Kim and Michael Ratz for 
useful discussions. 
KSC is supported in part by the Grant-in-Aid for
Scientific
Research No. 20$\cdot$08326 and 20540266 from the Ministry of
Education, Culture, Sports, Science and Technology of Japan.
P.V. would 
like to thank LMU Excellent for support.
This research was supported by the DFG cluster of excellence 
Origin and Structure of the Universe, the
European Union 6th framework program  MRTN-CT-2006-035863 ``UniverseNet'' 
and SFB-Transregio 33 "The Dark Universe" by Deutsche
Forschungsgemeinschaft (DFG).


\providecommand{\bysame}{\leavevmode\hbox to3em{\hrulefill}\thinspace}

\end{document}